# Relativistic aberration and null Doppler shift within the framework of superluminal and subluminal nondiffracting waves

Peeter Saari[1,2] and Ioannis M. Besieris[3]

[1] Institute of Physics, University of Tartu, W. Ostwaldi 1, 50411, Tartu, Estonia
[2] Estonian Academy of Sciences, Kohtu 6, 10130, Tallinn, Estonia
[3] The Bradley Department of Electrical and Computer Engineering, Virginia Polytechnic Institute and State University, Blacksburg, VA 24060, USA

E-mail: peeter.saari@ut.ee; besieris@vt.edu

**Abstract**

The relativistic aberration of a wavevector and the corresponding Doppler shift are examined in connection with superluminal and subluminal spatiotemporally localized pulsed optical waves. The requirement of a null Doppler shift is shown to give rise to a speed associated with the relativistic velocity composition law of a double (two-step) Lorentz transformation. The effects of such a transformation are examined both in terms of four-coordinates and in the spectral domain. It is established that a subluminal pulse reverses its direction. In addition to a change in direction, the propagation term of a superluminal pulse becomes negative. The aberration due to a double Lorentz transformation is examined in detail for propagation invariant superluminal waves (X wave, Bessel X wave), as well as intensity-invariant superluminal and subluminal waves. Detailed symmetry considerations are provided for the superluminal focus X wave and the subluminal MacKinnon wavepacket.

## 1. Introduction

The aberration of the wave vector due to the relative motion of reference frames and the corresponding Doppler shift in frequency are subjects not easily comprehended by students and are commonly associated with astrophysical and high-energy-physics fields. In particular, the treatment of the transverse Doppler effect and conditions of a null frequency shift depending on the choice of reference are not discussed in detail in textbooks.

The goals of the present study are, first, to consider these effects in connection with nondiffracting spatiotemporally localized pulsed optical waves (LWs) which not only can be treated as Lorentz-transformed versions of certain simple fields, but also are easily feasible in a laboratory of ultrafast optics. The second goal—which is of interest for the theory of LWs—is to reveal the symmetry properties of LWs with respect to certain double Lorentz transformations.

As a specific class of structured electromagnetic fields, LWs were mathematically discovered in the late 1980s, and since then a massive literature has been devoted to their theoretical and experimental study (see collective monographs [1, 2] and reviews [3-8]).

LWs are characterized by a specific type of space-time coupling. For all monochromatic plane-wave constituents of such pulsed waves in free space, there is a linear functional dependence between their temporal frequency (or wavenumber $k = \omega/c$) and the component $k_z$ of the wave vector in the direction of propagation of the pulses, rendering them propagation-invariant. This means that the spatial distribution of the pulse energy density does not change during propagation—it does not spread either in the lateral or in the longitudinal direction (or temporally). In reality, such a non-diffracting (non-spreading) propagation occurs over a large but still finite distance, because the aforementioned frequency-wavenumber functional dependence is not strict for practically realizable (finite-energy and finite-aperture) pulses.

The group velocity of LWs in empty space without the presence of any resonant medium not only can be smaller or equal to the vacuum speed of light $c$, but also can exceed $c$, i.e., it can be superluminal—in contradistinction with common superficial textbook treatments of the notion of group velocity of light pulses. This intriguing property has been widely discussed in the literature referred to above and has been experimentally verified by several groups [9-12] for cylindrically symmetric pulses. LWs that are localized in all three spatial dimensions and also temporally —which we shortly label as (3+1)D wavepackets—are in practice easily realizable only for certain types of superluminal and extremely subluminal pulses. At the same time, nondiffracting (2+1)D wave packets (pulsed light sheets), in which case the localization is sacrificed in one spatial dimension, can be flexibly generated by using spatial light modulators. Such technique has been recently introduced by Abouraddy's group at the University of Central Florida, resulting in the development of a wide program of study of such space-time wave packets (see, e.g., [8, 13, 14] and, particularly [15], where group velocities up to $30c$ were realized).

In our recent work [16], it was established that the velocity $v_e$ (normalized with respect to $c = 1$) at which the energy of LWs flows in the direction of propagation, is not equal to the propagation velocity of the pulse itself (i.e., to the normalized group velocity $v_g$). Instead, on the symmetry axis and/or on the locations of energy density maxima, these two quantities obey a simple relationship: specifically, $v_e = 2v_g \left(1 + v_g^2\right)^{-1}$. This is a physically content-rich result, because the right-hand-side expression arises from a two-step Lorentz transformation involving the normalized group velocity $v_g$. In section 2, where a brief introduction to the notion of X waves—the simplest representatives of superluminal LWs (or space-time wave packets) is given, it will be shown that this relationship arises naturally from the requirement of a null Doppler effect. In section 3, the double (two-step) Lorentz transformation of a general expression of LWs will be derived, first within the framework of four-coordinate transformations and then using transformations in the spectral domain. The symmetry properties of propagation-invariant subluminal and superluminal LWs under the double Lorentz transformations will be also considered there. Specific illustrative examples of the symmetry properties under the double Lorentz transformations will be given in section 4 for two typical propagation-invariant localized waves: the superluminal focus X wave and the subluminal MacKinnon wave packet, both of which have well-known closed-form analytical expressions. It should be noted that for the sake of simplicity the pulses are treated as scalar-valued fields, i.e., they may describe one component of an electromagnetic vector field. Nevertheless, the results can be extended to vector fields without loss of their validity, as was shown for the energy flow velocity in [16, 17]. In section 4 we also briefly discuss the differences in symmetry properties of sub- and superluminal LWs. Concluding remarks are given in section 5.

## 2. Aberration in two-step Lorentz transformations

### 2.1 Propagation-invariant X waves in terms of aberration

A primed reference frame is assumed to be moving along the $z$-axis with a speed $u$ with respect to an unprimed laboratory frame. Then, the wavevectors and frequency of a (3+1)D (three-dimensional + time) plane wave in free space transform as follows:

$$k_z' = \gamma(k_z - \beta k), \quad k_x' = k_x, \quad k_y' = k_y;$$
$$k' = \gamma(k - \beta k_z);$$
$$k' = \omega'/c, \quad k = \omega/c;$$
$$\gamma = 1/\sqrt{1-\beta^2}; \quad \beta = u/c, \quad u < c; \quad \beta = c/u, \quad u > c.$$
(1)

Let us first consider a monochromatic plane wave propagating at an angle $\theta$ with respect to the $z$-axis, as shown in figure 1. The angle $\theta$, called axicon angle, is defined by the relationship $k_z = k\cos\theta$. If we observe this wave in the primed reference frame which moves with a normalized speed $\beta = \cos\theta$, then according to Eq. (1) $k_z$ turns to zero, i.e., the wave propagates perpendicularly to the $z$-axis as shown in figure 1.

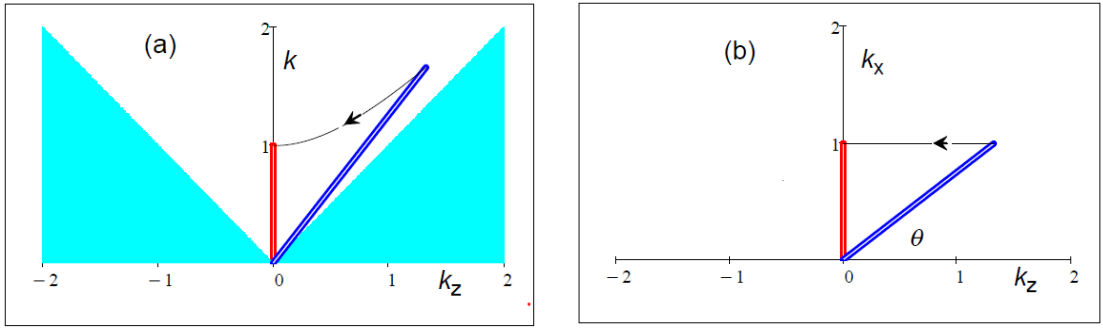

**Figure 1.** Lorentz transformation with $\beta = \cos\theta$ of a wavevector (if the thick lines are interpreted as components of the wavevector of a monochromatic plane wave) or of the support of a spectrum (if the thick lines are interpreted as projections of the support line of the spectrum of a wide-band plane wave

pulse). The arrows indicate the course of the transformation in the $(k, k_z)$-plane (a-panel) and in the $(k_x, k_z)$-plane (b-panel), respectively. The units are arbitrary (or of the order of $10^4$ cm$^{-1}$ in the optical regime). Shaded triangles in the panel (a) depict the regions outside the light cone which is forbidden because the dispersion relation prescribes the condition $k^2 \geq k_z^2$.

Figure 1 also depicts the frequency redshift as a result of the transverse Doppler effect: the smaller length of the wavevector in the primed reference frame, as indicated by the change of $k$-component in figure 1a, and the shortening of the vector in figure 1b.

In order to consider X waves, let us make two generalizations. First, we take a symmetrical pair of plane waves—the propagation direction of the first one lies on the $(x, z)$ plane and is inclined by an angle $+\theta$ with respect to the $z$ axis, and the second one by an angle $-\theta$ on the same plane. This does not change figure 1a because we are dealing with an analytic-signal representation of a field where negative frequencies are absent. But the vectors in figure 1b have mirrored counterparts with negative $k_x$ values. Second, let the waves be wide-band pulses. Then, the thick lines in figure 1 depict the supports of the spectra of the pulses, and the slope of the support line $k = k_z / \cos\theta$ gives group velocity (normalized with respect to $c = 1$) of the pulse propagating along the $z$-axis. In the primed reference frame the slope and the group velocity are infinitely large because the pulses are counterpropagating with respect to each other along the $x$-axis.

In the unprimed frame, the pulses form an X wave whose apex propagates without any spreading along the $z$-axis with a superluminal normalized (group) speed $v_g = 1/\beta$ (see figure 2). Practically realizable, i.e., finite-energy and finite-aperture pulses are non-spreading over a finite (although large) distance. In the spectral domain, this is reflected in a somewhat relaxed strictness of the functional bound $k = k_z / \cos\theta$ and, correspondingly, in the finite thickness of the support line.

In figure 2a, the apex is twice (intensity 4-fold) higher than that of the constituent pulses, while in the transverse direction the field is constant. However, if we make a generalization to a 3D setting, viz., to axisymmetric cylindrical waves, $k_x$ has to be replaced by the transverse wavenumber $k_\rho = \sqrt{k_x^2 + k_y^2}$ (see figure 1b) and the transverse dependence of the field acquires the profile of the ordinary Bessel function $J_0$ modulated by the function of the pulse envelope. Note that the relation $k = k_z / \cos\theta$, which determines the superluminal group velocity, remains the same. The field intensity along the generatrices of the double conical profile of the field, or along the "branches of X", decreases as $1/\rho$ (in contrast to the 2D case).

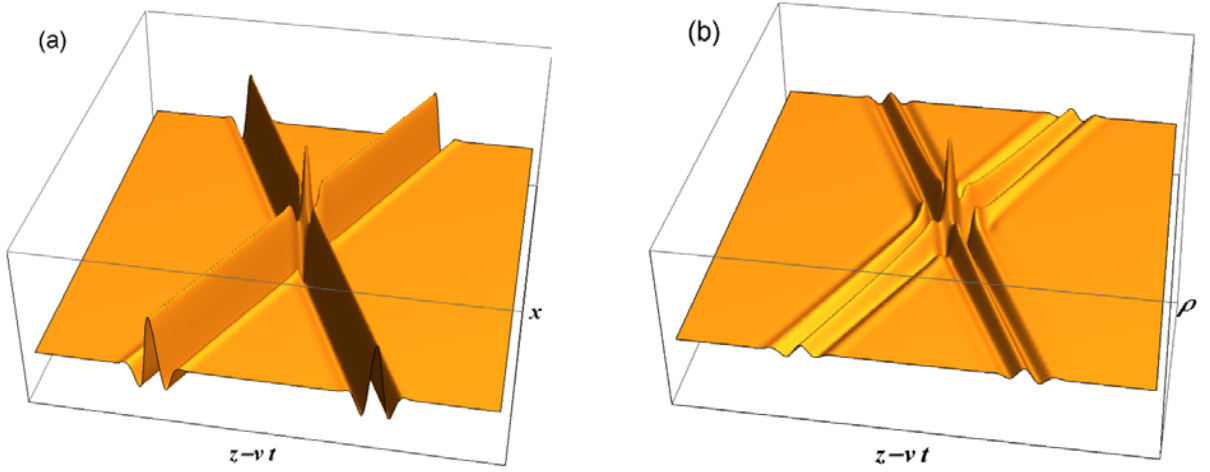

**Figure 2.** (a) X-wave pulse formed by two single-cycle cosine pulses (normalized group velocity $v_g = 1.11 = 1/\cos\theta$, i.e., $\beta = \cos\theta = 0.9$; full width of the pulse's Gaussian envelope at $1/e$-level is equal to the wavelength); (b) 3D Bessel-X pulse with the same parameters as in panel (a). The vertical axis indicates field strength normalized to peak value of the apex. The pulses propagate rigidly to the right as they depend on the coordinate $z$ and time $t$ only through the propagation variable $z - vt$, where $v = v_g c$ is the group velocity. The apex of the Bessel-X pulse moves along the $z$-axis where the radial coordinate $\rho = 0$. In order to better reveal the pulse profile, $\rho$ is shown as running over not only positive but also negative values; hence, due to the cylindrical symmetry, it represents any transverse coordinate.

Bessel-X pulses have been studied in a great number of papers (see reviews [1, 2, 6]), including their experimental realization in optics [10-12]. Although 2D X-type waves are poorly localized, they can be experimentally generated in a wide range of group velocities without encountering substantial technological obstacles, and they have become a subject of intensive study recently [8, 13-15]. For the present study, the most important conclusion from figure 1 and this introductory survey of X-waves is that they can be obtained through a Lorentz transformation from a pair of counterpropagating pulsed plane waves (in the 2D case), or from a radially collapsing and thereupon expanding cylindrical pulse (in the axisymmetric 3D case) [4, 6].

*2.2 The double Lorentz transformation*

The Doppler effect vanishes if the frequencies in the laboratory and moving frames are equal, i.e., $k' = k$. Given the relationship $k_z = k\cos\theta$, Eq. (1) leads to the condition $1 = \gamma(1 - \beta\cos\theta)$. The solution of the latter defines a new moving frame whose (normalized) speed relative to the unprimed (laboratory) frame is

$$\beta_e = \frac{2\cos\theta}{1+\cos^2\theta} = \frac{2\cos^{-1}\theta}{1+\cos^{-2}\theta} = \frac{2v_g}{1+v_g^2}.$$

(2)

As will be shown in detail in the next subsection, Eq. (2) defines the speed of a double (or two-step) Lorentz transformation from the laboratory reference to a second frame moving relative to the primed reference with speed $\beta = \cos\theta$, or with the relativistically doubled superluminal group velocity of a X-type pulse. It is thought-provoking that Eq. (2), as shown in [16, 17], also gives the on-axis energy flow speed $\beta_e c$ of (2+1)D and (3+1)D X-type pulses as well as of monochromatic Bessel beams $\exp(im\phi) J_m(k_\rho \rho) \exp[i(k_z z - kct)]$, with a radial wavenumber $k_\rho = \sqrt{k_x^2 + k_y^2}$.

With $\beta = \beta_e$ in the moving frame, one has

$$k_{z,e} = \gamma(k\cos\theta - \beta k)\big|_{\beta=\beta_e} = -k\cos\theta. \tag{3}$$

This result, a special case of the aberration phenomenon, means for X-type waves with propagation-invariant wavefunctions, that in a frame moving with speed $\beta_e$ along the $z$-axis their motion is *reversed*.

### 2.3 Relation to the relativistic velocity composition law

Consider two velocities, $u_{1,2}$ (not necessarily smaller than $c$) and the dimensionless quantities $\beta_{1,2} = u_{1,2}/c$. Sequential Lorentz transformations, first with the velocity $u_1$ and then with $u_2$, result in the velocity composition law

$$\beta_1 \oplus \beta_2 = \frac{\beta_1 + \beta_2}{1 + \beta_1 \beta_2} = \frac{\beta_1^{-1} + \beta_2^{-1}}{1 + \beta_1^{-1}\beta_2^{-1}}, \tag{4}$$

or, with dimensions,

$$u_1 \oplus u_2 = \frac{u_1 + u_2}{1 + \frac{u_1 u_2}{c^2}} = \frac{\frac{c^2}{u_1} + \frac{c^2}{u_2}}{1 + \frac{c^2}{u_1 u_2}}. \tag{5}$$

If $\beta_1 = \beta_2 = \beta = \cos\theta$, for example the normalized subluminal group speed of a pulsed Bessel beam, which is not propagation-invariant, or, for example the normalized superluminal group speed $1/\beta = 1/\cos\theta$ of a Bessel X pulse, a double Lorentz transformation with $\beta$ results in the expression

$$\beta_e = \beta \oplus \beta = \frac{2\beta}{1+\beta^2} = \frac{2\beta^{-1}}{1+\beta^{-2}}. \tag{6}$$

Eq. (6) arising from a double Lorentz transformation is identical to Eq. (2), derived from the requirement of a null Doppler effect with $\beta = \cos\theta$.

It can be shown that the following equalities also hold:

$$\beta \oplus (-\beta_e) = -\beta, \qquad (7a)$$

$$\beta^{-1} \oplus (-\beta_e) = -\beta^{-1}, \qquad (7b)$$

$$(-\beta) \oplus \beta_e = \beta.$$

It should be noted that (7a) expresses in terms of the composition of velocities the same result that Eq. (3) does in terms of the Lorentz-transformed $z$-component of the wavevector.

## 3. Transformation of subluminal and superluminal localized waves into a frame moving with velocity $\beta_e c$

*3.1 Transformation of 4-coordinates*

In the case of propagation-invariant localized waves (e.g., the X or Bessel-X pulses considered above), a normalized group velocity $v_g = 1/\beta$ enters into the propagation variable $z - v_g ct$ which expresses also the phase velocity. For intensity-invariant localized waves (e.g., the superluminal focus X wave (FXW) and the subluminal MacKinnon wavepacket [3, 4, 6]), the phase velocity is determined by the reciprocal of the normalized group velocity. Hence, the phase propagation variable is $z - \beta^{-1} ct$ for the MacKinnon pulse and $z - \beta ct$ for the FXW.

According to Eq. (4), the value of $\beta_e$ does not depend on whether it is determined through $\beta$ or $\beta^{-1}$. If $z$ and $ct$ are transformed according to the Lorentz transformations

$$z = \frac{z'' \pm \beta_e ct''}{\sqrt{1-\beta_e^2}}, \quad ct = \frac{ct'' \pm \beta_e z''}{\sqrt{1-\beta_e^2}}, \qquad (8)$$

one obtains

$$z - \beta ct \Rightarrow z + \beta ct,$$
$$z - \beta^{-1} ct \Rightarrow -(z + \beta^{-1} ct) \qquad (9)$$

if the positive signs are used in Eq. (8) corresponding to a double Lorentz boost in the direction of the pulse propagation in the laboratory frame. In Eq. (9) and from here on we omit the double primes on the right-hand side of transition expressions for brevity. It is seen, then, that a subluminal signal reverses its direction. In addition to a change in direction, the superluminal propagation variable becomes negative.

*3.2 Transformation of the spectral parameters.*

An expression for a localized solution to the scalar wave equation in free space is given by the spectral synthesis [4, 9, 16]

$$u(\rho,\phi,z,t) = e^{im\phi} \int_{-\infty}^{\infty} dk_z \int_{-\infty}^{\infty} dk\, S(k_z,k)\, J_m\left(\rho\sqrt{k^2-k_z^2}\right) \exp[i(k_z z - kct)];$$
$$S(k_z,k) = S_\rho\left(\sqrt{k^2-k_z^2}\right) \delta(k - v_g k_z - b) \Theta(k^2 - k_z^2), \tag{10}$$

where $J_m(\cdot)$ is the ordinary Bessel function of order $m$, $\Theta(\cdot)$ denotes the Heaviside unit step function and $S(\cdot)$ is a spectral distribution. To obtain an analytic signal, the factor $2\Theta(k)$ must be added. The $\delta$-function ensures non-spreading propagation with normalized group velocity $v_g$ by imposing a linear functional dependence between the wavenumber $k$ and the $z$-component of the wave vector. In contrast to the relationship used in section 2 and figure 1, the dependence is not a simple proportionality anymore due to the presence of the constant $b$.

The terms $k_z z - kct$, $dkdk_z$ and $k^2 - k_z^2$ in the integral above are Lorentz invariant. A Lorentz transformation of the argument of the Dirac delta function results in the expression

$$\delta(k - v_g k_z - b) \Rightarrow \frac{1}{|\gamma_2(1+v_g \beta_2)|} \delta\left(k - \frac{v_g + \beta_2}{1+v_g \beta_2} k_z - \frac{b}{\gamma_2(1+v_g \beta_2)}\right), \tag{11}$$

where $\gamma_2 = (1-\beta_2^2)^{-1/2}$ and $\beta_2$ is the speed of the moving frame normalized to $c=1$. The argument of the $\delta$-function in Eq. (11) indicates the transformed support line and, specifically, the coefficient of $k_z$ within the parenthesis in Eq. (11) is the Lorentz-transformed group speed. The last term is the additive constant in the moving frame. Thus, we have obtained the transformation rules

$$v_g \Rightarrow \frac{v_g + \beta_2}{1+v_g \beta_2}, \quad b \Rightarrow \frac{b}{\gamma_2(1+v_g \beta_2)}. \tag{12}$$

Next, we let the moving frame be the doubly primed frame, i.e., $\beta_2 = -\beta_e = -2v_g(1+v_g^2)^{-1}$ with $\gamma_2 = \gamma_e = (1-\beta_e^2)^{-1/2}$. It follows, then, from Eq. (12), with the equality $|\gamma_e(1-v_g \beta_e)| = 1$, that

$$\begin{aligned} v_g &\Rightarrow -v_g, \\ b &\Rightarrow -b, \text{ for } v_g > 1, \\ b &\Rightarrow b, \text{ for } |v_g| < 1. \end{aligned} \tag{13}$$

Thus, for both subluminal and superluminal pulses a double Lorentz transformation reverses the propagation direction. The sign of the parameter $b$ stays the same for subluminal waves but is reversed for superluminal ones.

## 4. Typical Subluminal and superluminal localized waves

*4.1 General expressions of subluminal and superluminal localized waves*

The spectral synthesis in Eq. (10) can be used to obtain the following forms of general subluminal and superluminal localized wave solutions to the scalar wave equation in free space:

$$u_{sub}(\rho,\phi,z,t) = e^{im\phi} \int_{-\infty}^{\infty} d\lambda F(\lambda) J_m\left(\frac{\rho}{\gamma}\sqrt{\gamma^4 b^2 - \lambda^2}\right) \exp[i\lambda(z-\beta ct)]$$
$$\times \exp\left[ib\gamma^2\beta\left(z-\frac{c}{\beta}t\right)\right], \tag{14a}$$

$$u_{sup}(\rho,\phi,z,t) = e^{im\phi} \int_{-\infty}^{\infty} d\lambda F(\lambda) J_m\left(\frac{\rho}{\gamma}\sqrt{\lambda^2 - \gamma^4 \beta b^2}\right) \exp\left[i\lambda\left(z-\frac{c}{\beta}t\right)\right]$$
$$\times \exp\left[ib\gamma^2\beta(z-\beta ct)\right]. \tag{14b}$$

Here, $\gamma = (1-\beta^2)^{-1/2}$, $\beta = v/c < 1$. The variable $\lambda$ is related to $k_z$ in Eq. (10) as follows: $\lambda = k_z - \gamma^2 \beta b$ in Eq. (14a) and $\lambda = k_z + \gamma^{-2}\beta b$ in Eq. (14b). Both integrals give rise to infinite-energy solutions. Finite-energy solutions can be derived by appropriate superpositions over the free parameter $b$. Integration over $\lambda$ in Eq. (14a) results in a localized wave with an envelope moving at the subluminal group speed $\beta c$ and it is modulated by a plane wave moving at the superluminal phase speed $c/\beta$. On the other hand, integration over $\lambda$ in Eq. (14b) results in a localized wave with an envelope moving at the superluminal group speed $c/\beta$ and modulated by a plane wave moving at the subluminal phase speed $\beta c$.

*4.2 Focus X wave (FXW)*

An axisymmetric localized wave with an envelope moving at a superluminal group speed $c/\beta$ is the FXW given as follows [4, 6]:

$$u(\rho,z,t) = \frac{\exp\left[-|b\beta|\gamma\sqrt{\rho^2 + \left(a - i\beta\gamma\left(z-\frac{c}{\beta}t\right)\right)^2}\right]}{\sqrt{\rho^2 + \left(a - i\beta\gamma\left(z-\frac{c}{\beta}t\right)\right)^2}} \exp\left[-ib\beta\gamma^2(z-\beta ct)\right]. \tag{15}$$

The doubly Lorentz-transformed signal is given by

$$u_{DLT}(\rho,z,t) = \frac{\exp\left[-|b\beta|\gamma\sqrt{\rho^2 + \left(a + i\beta\gamma\left(z + \frac{c}{\beta}t\right)\right)^2}\right]}{\sqrt{\rho^2 + \left(a + i\beta\gamma\left(z + \frac{c}{\beta}t\right)\right)^2}} \exp\left[-ib\beta\gamma^2(z + \beta ct)\right]. \quad (16)$$

Again, this follows from both the transformation of four-coordinates [see Eq. (9)] and the transformation of spectral parameters [see Eq. (13)]. The similar expressions in Eqs. (15) and (16) are related as follows: $u_{DLT}(\rho,z,t;\beta,b) = u^*(\rho,z,-t;\beta,-b)$ and $u_{DLT}(\rho,z,t;\beta,b) = u(\rho,z,t;-\beta,-b)$. We see that the symmetry relations between the FXW and its transformed version are more complicated than simply time reversal and complex conjugation. Should the latter have been the case, the real part of Eq. (16) would be simply the time reversed replica of the real-valued original FXW and polarity reversal would be added for the imaginary part (both parts being possible real solutions to the wave equation). Responsible for the complications is the parameter $b$ which changes the simple proportionality relation between wavenumber $k$ and the $z$-component of the wave vector considered in section 2 to a more general linear dependence. As can be seen in figure 3, pulses given by Eqs. (15) and (16) with identical parameters, differ substantially. Specifically, the wavelength of the carrier of the transformed FXW for the given parameters is 4 times shorter and, correspondingly, its frequency 4 times higher than those of the original FXW. Mathematically, the differences arise from the symmetry relations between Eqs. (15) and (16). To reveal the physical reasons and study the differences in more detail, let us first look at the transformation of the supports of the spectrum of FXW in figure 4.

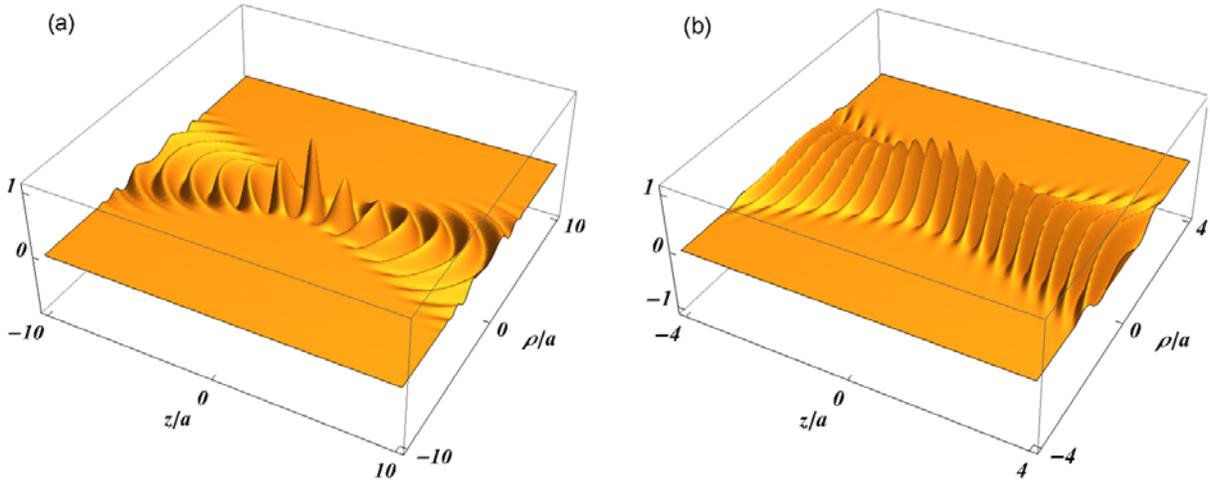

Figure 3. Real part of Eq. (15)—(a), and Eq. (16)—(b), at the instant $t = 0$. Parameters: $\beta = 0.6$ and $b = 10/a$. With the choice $a = 1\mu m$, the carrier wavelength of the transformed pulse (b) is $0.4\mu m$, i.e., falls into the visible region. Note the different scales of coordinates in (a) and (b).

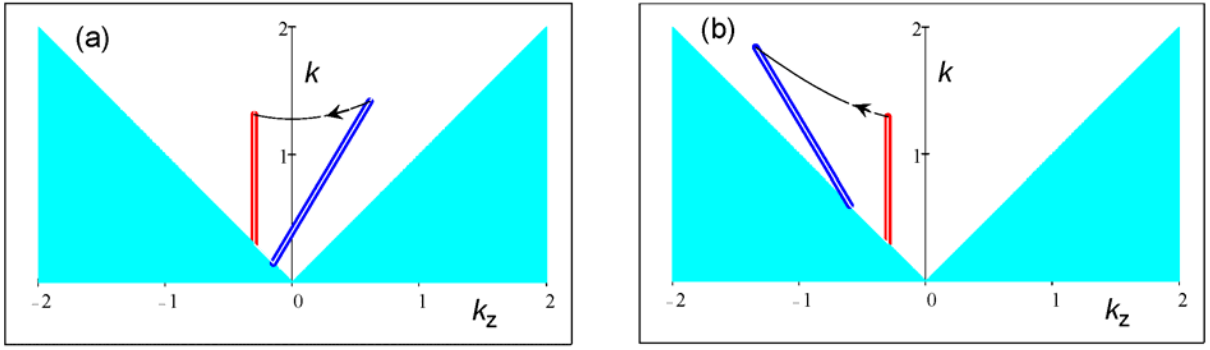

Figure 4. Lorentz transformations of the support of the spectrum of the superluminal FXW with parameters $b = 0.4a^{-1}$ and $\beta = 0.6$. (a) To the primed reference frame; (b) the second step resulting in the spectral support of $u_{DLT}$. The scales are in units of $a^{-1}$. See also the caption of figure 1.

The result of the first Lorentz transformation is the vertical support line at $k_{z0} = -b\beta\gamma = -0.3$. It means that in the primed frame one observes a radially collapsing and thereupon expanding cylindrical pulse, which—in contrast to the pulse considered in section 2—is modulated sinusoidally along the $z$-axis with spatial frequency $k_{z0}$ [6]. Since the spectrum decays exponentially with the wavenumber as $\exp(-ak)$, its support stretches to infinity, but on the figure its length has been cut to value 1, i.e., at the point where the strength of the spectrum drops below level $1/e$. Figure 4 shows the reason why $u_{DLT}(.)$ is not simply a time-reversed copy of $u(.)$: their spectra do not possess mirror symmetry with respect to each other as was the case of X waves in section 2; instead, they differ substantially and, in particular, the mean wavenumber increases due to the double Lorentz transformation. As a result of the latter the parameters of the support line obviously change in accordance with Eq. (13). All this is caused by the additive constant $b$ in the argument of the $\delta$-function already in the general expression Eq.(10). Also, this constant spoils the simple relation in Eq. (3) because the condition of a null Doppler effect and the axicon angle $\theta$ acquire a dependence on the wavenumber.

Eq. (15) at $\rho = 0$ yields the following expressions for its phase

$$\Psi = -b\beta\gamma^2(1-\beta)(z+ct)+\Phi, \text{ if } b > 0, \beta > 0 \text{ or } b < 0, \beta < 0; \tag{17a}$$

$$\Psi = -b\beta\gamma^2(1+\beta)(z-ct)+\Phi, \text{ if } b < 0, \beta > 0 \text{ or } b > 0, \beta < 0. \tag{17b}$$

Here, $\Phi = \arctan[\beta\gamma(z-\beta^{-1}ct)/a]$ is a Gouy-type contribution to the phase, which comes from the denominator of Eq. (15). Bearing in mind our parameter values, Eq. (17a) is the case. It follows that contrary to what one might think looking at Eq. (15), the carrier moves in the *negative* direction of the $z$-axis *with velocity c*. Using the symmetry relation $u_{DLT}(\rho,z,t;\beta,b) = u(\rho,z,t;-\beta,-b)$, Eq. (17a) allows one to immediately deduce the on-axis behaviour of the transformed field given by Eq. (16): the movement is the same but its wavenumber or frequency is $(1+\beta)/(1-\beta)$ higher, which makes it 4 times with our parameters, in full accord with the oscillation periods (far from the pulse maximum where the period is affected by the Gouy phase jump) in figures 5a and 5b.

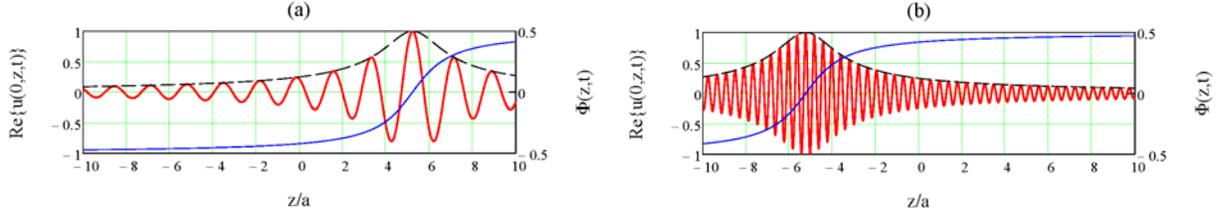

Figure 5. On-axis spatial behaviour of the real part (oscillating curve), amplitude (dashed curve), and the Gouy-type contribution to the phase of (a) $u(.)$ and (b) $u_{DLT}(.)$ at the time instant $ct = \pi a$. The right-hand scale is in units of $\pi$.

In order to get a transformed version of the FXW which would have the same spectrum as the original, one has to make replacement $\beta \Rightarrow -\beta$ in Eq. (15) but leave the sign of $b$ *unchanged*. This way we get a new solution of the wave equation which is related to the original as

$$u^{(2)}(\rho,z,t) = u^*(\rho,z,-t), \tag{18}$$

i.e., it is simply a backward-propagating version on the initial FXW (the imaginary part being additionally of opposite polarity).

### 4.3 MacKinnon subluminal wavepacket

A simple axisymmetric subluminal localized wave is the MacKinnon wavepacket given by Eq. (19) [4, 6, 16].

$$u(\rho,z,t) = \frac{\sin\left[|b|\gamma\sqrt{\rho^2 + \gamma^2(z-\beta ct)^2}\right]}{\sqrt{\rho^2 + \gamma^2(z-\beta ct)^2}} \exp\left[ib\gamma^2\beta\left(z-\frac{c}{\beta}t\right)\right]. \tag{19}$$

It can be considered as a spherical standing monochromatic wave, viewed from a reference frame moving with the subluminal speed $\beta$, whose spherical intensity profile has compressed to a spheroidal one due to the Lorentz contraction in the direction of the propagation axis $z$ (see figure 6).

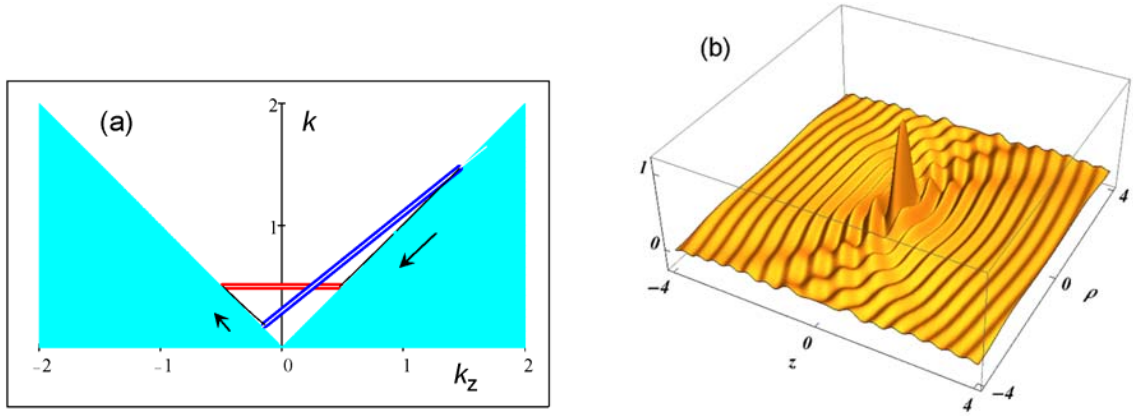

Figure 6. (a) Lorentz transformation of the support of the spectrum of the subluminal MacKinnon pulse with parameters $\beta = 0.8$ and $b = 0.3$ to the primed reference frame (unit of wavenumbers is $4\pi\lambda^{-1}$, where $\lambda$ is wavelength of the standing wave in the primed frame. (b) The real part of Eq. (19) with parameters $\beta = 0.8$ and $b = 3.77\lambda^{-1}$ at the instant $t = 0$; the units of the longitudinal and transverse axes are $\lambda$. If $\lambda = 1\mu m$, the field oscillations fall into the visible region and, particularly, their wavelength along the axis $z$ is $0.45\mu m$.

Either the transformation of four-coordinates [see Eq. (9)] or the transformation of spectral parameters [see Eq. (13)] give the doubly Lorentz-transformed version of Eq. (19):

$$u_{DLT}(\rho,z,t) = \frac{\sin\left[|b|\gamma\sqrt{\rho^2 + \gamma^2(z+\beta ct)^2}\right]}{\sqrt{\rho^2 + \gamma^2(z+\beta ct)^2}} \exp\left[-ib\gamma^2\beta\left(z+\frac{c}{\beta}t\right)\right]. \qquad (20)$$

The expressions in Eqs. (19) and (20) are related as follows: $u_{DLT}(\rho,z,t) = u^*(\rho,z,-t)$ and $u_{DLT}(\rho,z,t,\beta) = u(\rho,z,t;-\beta)$. Consequently, unlike the FXW, a real-valued Mackinnon wavepacket transforms to a backward propagating one. The reason why this simple symmetry relation holds in the case of the MacKinnon pulse, despite the presence of the constant $b$ in the support line of its spectrum, is the following. The first step of the double Lorentz transformation changes the spectral support to a horizontal line as depicted in figure 6a, which corresponds to a directionally uniform spatial spectrum of the monochromatic standing wave (with $k = 0.5$ in figure 6a). Since the spectrum is constant over its support in the form of a spherical surface in the $k$-space of the primed frame, the second step of the double Lorentz transformation results in a spectrum with its support that is mirrored as $z \to -z$ from the initial one. Therefore, the MacKinnon pulse as a special case of a subluminal LW is temporally symmetrical under the double Lorentz transformation. It can be shown that any finite-energy version of the MacKinnon pulse, obtainable by integration of Eq. (19) with respect to the parameter $b$ over a spectral distribution function, possesses the same temporal-directional symmetry.

## 5. Concluding Remarks

It was mentioned in section 1 (and also in subsections 2.1 and 4.2) that spatiotemporally localized pulsed optical waves (LWs) can be treated as Lorentz-transformed versions of certain simple fields. Let us make it clear in connection with the FXW and the MacKinnon wavepacket. The former arises from the wavelet

$$v_{\text{sup}}(\rho,z,t) = \frac{\exp\left[-ikz - |k|\sqrt{\rho^2 + (a+ict)^2}\right]}{\sqrt{\rho^2 + (a+ict)^2}}, \quad k = b\beta\gamma \tag{21}$$

by means of the inverse Lorentz transformation

$$z \to \frac{z - \beta ct}{\sqrt{1-\beta^2}}, \quad ct \to \frac{ct - \beta z}{\sqrt{1-\beta^2}}, \tag{22}$$

and the latter from the monochromatic spherically symmetric expression

$$v_{\text{sub}}(\rho,z,t) = e^{-ickt} \frac{\sin\left(|k|\sqrt{\rho^2 + z^2}\right)}{\sqrt{\rho^2 + z^2}}, \quad k = b\gamma, \tag{23}$$

again, by means of the inverse Lorentz transformation in Eq. (22).

A direct Lorentz transformation brings the FXW $u$ in Eq. (15) to the form $v_{\text{sup}}$ in Eq. (21) and an additional direct Lorentz transformation results in $u_{DLT}$ in Eq. (16). Similarly, a direct Lorentz transformation brings the MacKinnon wavepacket $u$ in Eq. (19) to the form $v_{\text{sup}}$ in Eq. (23) and an additional direct Lorentz transformation results in $u_{DLT}$ in Eq. (20).

The symmetry relationships arising from a double Lorentz transformation were discussed in detail in connection with the infinite-energy intensity-invariant FXW and the MacKinnon pulse. Corresponding finite-energy subluminal and superluminal localized waves can be derived using appropriate superpositions over the free parameter $b$. All the symmetry properties are preserved in this case.

Finally, a note on the physical meaning of superluminality of finite-energy localized waves is appropriate. The presence of a superluminal speed does not contradict relativity. The pulse moves superluminally with almost no distortion up to a certain distance $z_d$, which is determined by the geometry (aperture size and axicon angle) and then it slows down to a luminal speed $c$, with significant accompanying distortion. Although the peak of the pulse does move superluminally up to $z_d$, it is not causally related at two distinct ranges $z_1, z_2 \in [0, z_d)$. Thus, no information can be transferred superluminally from $z_1$ to $z_2$.